\documentclass[aps,prc,showpacs,onecolumn,preprintnumbers,superscriptaddress,nofootinbib,floatfix,10pt]{revtex4-1}
\usepackage{graphicx}
\usepackage{amsmath,amssymb}
\usepackage{hyperref}
\usepackage{physics}

\begin{document}

\title{Investigation of the $\bar{K}$--$^{6}$Li Interaction 
and the Search for the $\Lambda(1405)$ Resonance}

\author{Ahmad Naderi Beni}
\affiliation{Department of Basic Sciences, Technical and 
Vocational University (TVU), Tehran, Iran}
\author{Sajjad Marri}
\email{marri@iut.ac.ir}
\affiliation{Department of Physics, Isfahan University 
of Technology, Isfahan 84156-83111, Iran}
\date{\today}

\begin{abstract}
We investigate the interaction of an antikaon ($K^-$) with the $^{6}$Li nucleus, 
described as an $\alpha + d$ cluster system. The study aims to explore the formation 
of the $\Lambda(1405)$ resonance through the $K^-d$ subsystem in the presence of a 
spectator $\alpha$ particle. In the absence of dedicated experimental data for this 
reaction, particular attention is given to providing quantitative predictions for 
the manifestation of the $\Lambda(1405)$ structure in low-energy $\bar{K}N$ dynamics 
within a light nuclear environment. Employing different models of the 
$\bar{K}N-\pi\Sigma$ interaction, we calculate the $\pi\Sigma n$ invariant-mass 
spectra and the $\alpha$-particle missing-mass spectra, thereby identifying robust 
features of the $\Lambda(1405)$ signal and offering guidance for future experimental 
investigations.
\end{abstract}

\maketitle
\section{Introduction}
Understanding the interaction between an antikaon and a nucleon remains a 
fundamental challenge in contemporary hadron physics. The $\bar{K}N$ system 
provides a unique window into the nonperturbative dynamics of quantum 
chromodynamics (QCD) in the strangeness sector, where the interplay between 
spontaneous chiral symmetry breaking and strong coupled-channel effects gives 
rise to rich phenomenology. One of the most intriguing manifestations of this 
dynamics is the appearance of the $\Lambda(1405)$ resonance, located just 
below the $\bar{K}N$ threshold. This state, often interpreted as a quasi-bound 
$\bar{K}N$ system embedded in the $\pi\Sigma$ continuum, cannot be easily 
explained within the simple three-quark framework of conventional baryon 
spectroscopy \cite{p1,p2,p3,p4,p5,p6,p7,c1,c2,c3,c4,c5,c55}.  

Theoretical models describing the $\bar{K}N$ interaction generally fall 
into two broad categories. Phenomenological approaches adjust potential 
parameters to reproduce low-energy scattering data and kaonic atom 
observables \cite{p1,p2,p3,p4,p5,p6,p7}. In contrast, chiral 
SU(3) dynamics derives the interaction kernels systematically from 
effective field theory consistent with QCD symmetries~\cite{c1,c2,c3,c4,c5,c55}. 
Although both frameworks reproduce experimental data around the $\bar{K}N$ 
threshold, their predictions deviate considerably when extrapolated to 
subthreshold energies, where the $\Lambda(1405)$ dominates~\cite{l1}. 
Precise experimental input such as $K^-p$ scattering data and kaonic 
hydrogen measurements~\cite{l2,l3,l4,l5,l6} strongly constrain 
the models, yet direct information below the $\bar{K}N$ threshold remains 
limited. Consequently, reactions producing $\pi\Sigma$ final states, in 
which the $\Lambda(1405)$ appears as a pronounced structure, play a 
crucial role in determining the subthreshold behavior of the $\bar{K}N$ 
amplitude.

The existence of the $\Lambda(1405)$ resonance was first predicted 
by Dalitz and Tuan~\cite{p1,p2} based on the analysis of $\bar{K}N$ 
scattering lengths, and later confirmed experimentally through the 
observation of the $\pi\Sigma$ invariant mass distribution in 
$K^-p \to \pi\pi\pi\Sigma$ reactions~\cite{l7}. Recent experimental 
developments have greatly improved our understanding of this resonance. 
The photoproduction experiments by the CLAS Collaboration at Jefferson 
Laboratory~\cite{l8,l9}, the LEPS measurements at SPring-8 \cite{m1,m2}, 
and studies by HADES~\cite{m3}, E31 at J-PARC \cite{m4,e31}, and KLOE 
at DA$\Phi$NE~\cite{m5}, have provided high-precision $\pi\Sigma$ spectra. 
Analyses of these results suggest that the $\Lambda(1405)$ may possess 
a two-pole structure \cite{m6,m7,m8}, reflecting its complex origin from 
the interplay of $\bar{K}N$ and $\pi\Sigma$ channels. 

The absorption of slow or stopped kaons in nuclei provides an effective 
environment for investigating $\bar{K}N$ interactions below threshold. 
When a kaon is captured by a bound proton, it can form the $\Lambda(1405)$ 
as an intermediate state, which subsequently decays into $\pi\Sigma$ 
channels. Thus, the invariant mass and momentum distributions of the 
emitted particles carry direct information on the $\bar{K}N$ amplitude 
and its in-medium modification. Such processes can also act as a doorway 
to the formation of kaonic nuclear clusters.   

In this context, the reaction \(\mathrm{A}(K^-,\pi\Sigma)\mathrm{A}'\) 
on light nuclei such as \(d\), \(^3\mathrm{He}\), and \(^4\mathrm{He}\) 
was investigated, and the resulting \(\pi\Sigma\) spectra exhibited 
features consistent with the formation of the \(\Lambda(1405)\) resonance 
in the nuclear medium~\cite{n1,n2,t2,t3,t4,t5,t6,t7,t8,t9}. In addition, 
measurements of pion momentum spectra from stopped-kaon absorption on 
various light nuclear targets, including \(^6\mathrm{Li}\), \(^7\mathrm{Li}\), 
\(^9\mathrm{Be}\), \(^{12}\mathrm{C}\), and \(^{16}\mathrm{O}\), provided 
further experimental information on antikaon--nucleus interactions~\cite{n3}.

Motivated by previous studies on kaon absorption in light nuclei, the 
present work focuses on the $K^- + {}^6\mathrm{Li}$ reaction, as 
investigated by the FINUDA experiment~\cite{n3,nn3}. The $^6\mathrm{Li}$ 
nucleus is modeled as an $\alpha + d$ cluster system, with the $\alpha$ 
particle acting as a spectator while the main dynamics occur in the $K^- d$ 
subsystem. In the absence of dedicated experimental data for the $\pi\Sigma n$ 
invariant-mass and $\alpha$-particle missing-mass spectra in this reaction, 
we provide quantitative predictions for the manifestation of the $\Lambda(1405)$ 
resonance within the light nuclear environment. This approach allows us to 
examine the robustness of the $\Lambda(1405)$ signal and to identify characteristic 
features of the $\bar{K}N$--$\pi\Sigma$ coupled-channel dynamics, which can 
serve as benchmarks for future experimental investigations. Furthermore, our 
calculations explore the dependence of the results on different $\bar{K}N$ 
interaction models, including phenomenological (SIDD1, SIDD2) and chiral approaches.

The paper is organized as follows. In Section~\ref{formula}, we outline the 
formalism used to calculate the transition amplitude for the $K^- +{}^6\mathrm{Li}$ 
reaction. Sections~\ref{put} and \ref{result} present the input interactions 
and discusses the resulting alpha momentum spectra and $\pi\Sigma{n}$ mass 
spectra, respectively. The Section~\ref{conc} summarizes the main conclusions 
of this work.
\section{Faddeev formalism for $K^- + {}^6\mathrm{Li}$ reaction}
\label{formula}
To investigate $K^- +{}^6\mathrm{Li}$ reaction, the \(^{6}\mathrm{Li}\) nucleus 
is described within a cluster framework as a \(d\)--\(\alpha\) system. Furthermore, 
the deuteron cluster is explicitly treated as a two-body system composed of a 
neutron and a proton, while no internal structure is assigned to the \(\alpha\) 
particle. Owing to the lack of reliable information on the antikaon--\(\alpha\) 
interaction, the \(\alpha\) particleis assumed to act as a spectator in the 
present calculation.

Within this framework, we focus on the antikaon--deuteron interaction in 
the presence of an \(\alpha\) particle. Although the full reaction dynamics 
corresponds to a four-body system, \(K^- + n + p + \alpha\), the spectator 
assumption allows us to reduce the problem to an effective three-body system 
involving \(K^-\), \(n\), and \(p\), with the influence of the \(\alpha\) 
particle entering only through overall kinematics. This approximation 
significantly simplifies the theoretical treatment while retaining the 
essential physics of the dominant absorption mechanism.

To calculate the \(\pi\Sigma n\) invariant-mass spectra for the reaction 
\(K^- + {}^{6}\mathrm{Li} \rightarrow \pi\Sigma n + \alpha\), we formulate 
the Faddeev equations for the \(\bar{K}NN\) subsystem with total spin \(s=1\) 
and isospin \(I=1/2\). We then evaluate the scattering amplitude for the 
\(K^- + d \rightarrow \pi + \Sigma + n\) reaction in the presence of a 
spectator \(\alpha\) particle. The observables associated with these 
subsystems are obtained by solving the corresponding three-body Faddeev 
equations. For a system composed of three particles \(i\), \(j\), and \(k\), 
the three-body Faddeev equations in the Alt--Grassberger--Sandhas (AGS) form 
can be written as~\cite{n5,nn5,nn6}:
\begin{widetext}
\begin{equation}
\mathcal{K}_{ij,I_{i} I_{j}}^{\alpha\beta}=
(1-\delta_{ij})\mathcal{M}_{ij,I_{i} I_{j}}^{\alpha}
+\sum_{k\neq i,I_{k};\gamma}\mathcal{M}_{ik,I_i I_k}^{\alpha} \,
\tau_{kI_k}^{\alpha\gamma} \, \mathcal{K}_{kj,I_k I_j}^{\gamma\beta}, 
\hspace{0.5cm} 
\qquad \alpha,\beta \,\mathrm{and} \, 
\gamma \in \{\bar{K}N_{1}N_{2},\pi\Sigma{N}_{2},\pi{N}_{1}\Sigma\},
\label{eeq1}
\end{equation}
\end{widetext}

Here, \(\mathcal{K}_{ij,I_i I_j}^{\alpha\beta}\) denotes the transition operators, 
\(\mathcal{M}_{ij,I_i I_j}^{\alpha}\) represents the corresponding driving terms, 
and \(\tau_{k I_k}^{\alpha\beta}\) stands for the two-body \(T\)-matrices embedded 
in the three-body system. The \(\bar{K}N\) interaction is dynamically coupled to 
the \(\pi\Sigma\) channel, while the \(\pi\Lambda\) channel is taken into account 
effectively through the employed interaction model. Consequently, the \(\pi\Sigma N\) 
configurations are explicitly included in the calculation, and the Faddeev equations 
in Eq.~(\ref{eeq1}) are extended to incorporate the \(\bar{K}N\text{--}\pi\Sigma\) 
coupled-channel dynamics.

In the present formulation, the total isospin of the interacting particles uniquely 
fixes the spin quantum numbers of the system. Consequently, the baryon spin degrees 
of freedom do not appear explicitly, and all operators are labeled solely by isospin 
indices. For the \(K^- d\) system, the spin wave function is symmetric; therefore, 
the corresponding operators in the isospin basis must be antisymmetric in order to 
satisfy the overall fermionic antisymmetry.

For an efficient solution of the three-body equations, it is advantageous to express 
both the three-body transition amplitudes and the driving terms in a separable form. 
This representation reduces the integral equations in Eq.~(\ref{eeq1}) to a homogeneous 
system, which significantly simplifies their numerical treatment. In the present work, 
we adopt the energy-dependent pole expansion (EDPE) method, following the formulation 
described in Refs. \cite{n6,n7,n8}. Within this framework, the separable representation 
of the Faddeev transition amplitudes can be written as:
\begin{equation}
\mathcal{K}_{ij,I_i I_j}^{\alpha\beta}(q,q';\epsilon)=
\sum_{\mu{,}\nu}^{N_{r}} 
u^{\alpha}_{\mu,iI_{i}}(q,\epsilon) 
\theta^{\alpha\beta;I_{i}I_{j}}_{ij;\mu\nu}(\epsilon) 
u^{\beta}_{\nu,jI_{j}}(q',\epsilon),
\label{eeq2}
\end{equation} 
where \(u^{\alpha}_{\mu,i I_i}\) denotes the form factors associated with the 
interacting subsystems, while \(\theta^{\alpha\beta; I_i I_j}_{ij;\mu\nu}(\epsilon)\) 
represents the elements of the coupling matrix that encode the interaction dynamics 
between different three-body partitions. The variables \(q\) and \(q'\) correspond 
to the momenta of the spectator particle in channels \(i\) and \(j\) of the three-body 
subsystem, respectively.

The eigenfunctions \(u^{\alpha}_{\mu,i I_i}(q,\epsilon)\) appearing in
Eq.~(\ref{eeq2}) are obtained by solving the corresponding homogeneous
Faddeev equations.
\begin{equation}
\begin{split}
& u^{\alpha}_{\mu,iI_{i}}(q,\epsilon)=
\frac{1}{\lambda_{\mu}}
\sum\limits_{j\neq{i},I_{j};\beta}\int 
\mathcal{M}^{\alpha}_{ij,I_{i}I_{j}}(q,q';\epsilon)\, 
\tau^{\alpha\beta}_{jI_{j}} 
\big(\epsilon \big) \\
& \hspace{1.6cm}\times 
u^{\beta}_{\mu,jI_{j}}(q',\epsilon)d^{3}q^{\prime},
\end{split}
\label{eeq3}
\end{equation}

By solving Eq.~(\ref{eeq3}), one can determine the possible binding energies 
\(B\) of the three-body \(\bar{K}NN\) system, along with the corresponding 
form factors \(u^{\alpha}_{\mu,i I_i}(q,B)\) and eigenvalues \(\lambda_{\mu}\) 
evaluated at \(\epsilon = B\). The binding energy and width of the \(\bar{K}NN\) 
subsystem were obtained by solving the single-channel Faddeev equations for 
the \(K^- d\) system. Using the SIDD1 potential for the antikaon--nucleon 
interaction, the binding energy is \(B = 1.9~\mathrm{MeV}\) with a width 
\(\Gamma = 70.8~\mathrm{MeV}\), while for the SIDD2 potential, the corresponding 
values are \(B = 5.6~\mathrm{MeV}\) and \(\Gamma = 63.4~\mathrm{MeV}\). These 
results are in good agreement with those reported in Ref.~\cite{nn8}, highlighting 
the sensitivity of the \(\bar{K}NN\) cluster properties to the choice of the 
\(\bar{K}N\) interaction model. The form factors are normalized according to the 
condition

\begin{equation}
\sum_{i;\alpha\beta} \int u^{\alpha}_{\mu,iI_{i}}(q,B) 
\tau^{\alpha\beta}_{iI_{i}} \big(q,B\big) 
u^{\beta}_{\nu,iI_{i}}(q,B)d^{3}q =-\delta_{\mu\nu}.
\label{eeq4}
\end{equation}

In Eq.~(\ref{eeq3}), the form factors are defined at a fixed energy 
\(\epsilon_{\sigma} = B_{\sigma}\), corresponding to the binding energy of the 
three-body system. In order to extend the applicability of the eigenfunctions 
\(u^{\sigma}_{\mu,i I_i}\) over the full energy and momentum range, an 
extrapolation procedure is performed by
\begin{equation}
\begin{split}
& u^{\alpha}_{\mu,iI_{i}}(q,\epsilon)=
\frac{1}{\lambda_{\mu}}
\sum\limits_{j\neq{i},I_{j};\beta}\int
\mathcal{M}^{\alpha}_{ij,I_{i}I_{j}}(q,q';\epsilon)\, \\
& \hspace{3cm}\times \tau^{\alpha\beta}_{jI_{j}} 
\big(q',B\big) 
u^{\beta}_{\mu,jI_{j}}(q',B)
d^{3}q^{\prime}.
\end{split}
\label{eeq5}
\end{equation}

After determining the eigenfunctions \( u^{\sigma}_{\mu,iI_{i}}(q,\epsilon_{\sigma}) \), 
one can define the effective EDPE propagators \( \theta(\epsilon_{\sigma}) \) 
in Eq.~(\ref{eeq2}) by
\begin{equation}
\begin{split}
& \big((\theta(\epsilon))^{-1}\big)_{\mu\nu}
=\sum_{iI_{i};\alpha\beta}\int
\big[u^{\alpha}_{\mu,iI_{i}}(q,B)
\tau^{\alpha\beta}_{iI_{i}} \big(q,B\big) \\
& \hspace{2.cm} -u^{\alpha}_{\mu,iI_{i}}(q,\epsilon)
\tau^{\alpha\beta}_{iI_{i}}\big({q,\epsilon}\big)
\big] u^{\beta}_{\nu,iI_{i}}(q,\epsilon)d^{3}q.
\end{split}
\label{eeq6}
\end{equation}

Based on Eq.~(\ref{eeq6}), the Faddeev indices ($i,j$ and $k$) and isospin 
indices ($I_{i}$ and $I_{j}$) of the $\theta(\epsilon)$
-functions are unnecessary and could be omitted. Therefore, we have
\begin{equation}
\theta^{I_{i}I_{j}}_{ij;\mu\nu}(\epsilon) = 
\theta_{\mu\nu}(\epsilon).
\label{eeq7}
\end{equation} 

The transition from the initial \(K^- + {}^{6}\mathrm{Li}\) state to the final 
\(\alpha + \pi\Sigma n\) channel proceeds through a sequence of nontrivial 
reaction mechanisms. In the first step, the system may evolve from 
\(K^- + {}^{6}\mathrm{Li}\) into an intermediate configuration consisting of 
an \(\alpha\) particle and a \((\bar{K}NN)_{s=1}\) subsystem, as illustrated 
in Fig.~\ref{fig1}. Alternatively, the reaction can proceed directly to the 
\(\alpha + \pi\Sigma n\) final state. At this stage, the dynamics corresponds 
to a genuine four-body system involving an antikaon interacting with an 
\(\alpha\)--\(np\) cluster.

In principle, a rigorous treatment of this process would require solving the 
inhomogeneous four-body Faddeev equations in order to determine the transition 
amplitude. However, due to the lack of reliable information on the kaon--\(\alpha\) 
interaction, we adopt a three-body approximation in the present work. Specifically, 
we assume that the reaction proceeds predominantly via the intermediate 
\(\alpha + (\bar{K}NN)_{s=1}\) state, in which the \(\alpha\) particle acts as a
spectator. The subsequent decay of the \((\bar{K}NN)_{s=1}\) subsystem in the 
presence of the spectator \(\alpha\) leads to the \(\pi\Sigma n\) final state.
Therefore, the scattering amplitude ($T$) for the 
$K^{-} + \mathrm{^{6}Li} \rightarrow  \alpha + (\pi\Sigma n)$ reaction channel can 
be defined by~\cite{n9}
\begin{widetext}
\begin{align}
  &T_{K^{-}+\mathrm{^{6}Li}\rightarrow \alpha+\pi\Sigma n}(\bm{p}_\alpha,\bm{q}_n,\bm{P}_{\bar{K}},W) 
  = \sum_{I;\mu\nu} \mathcal{A} \, \theta_{\mu\nu} (W-p_{\alpha},p_{\alpha}) \nonumber\\
 &\times\Big[\langle \pi\Sigma n|[[\pi\otimes \Sigma]_{I}\otimes N]_\Gamma\rangle~
 g_{\pi\Sigma}^{(I)}(k_n)~\tau_{\pi\Sigma-\bar{K}N}^{(I)}(W-E_\alpha(p_{\alpha})-E_{n}(\bm{q}_n))
 ~(u^{1}_{\nu,2I}(q,W-E_{\alpha})-u^{1}_{\nu,3I}(q_{n},W-E_{\alpha})) \nonumber\\
 & +\langle \pi\Sigma n|[[\pi\otimes \Sigma]_{I}\otimes N]_\Gamma\rangle~
 g_{\pi\Sigma}^{(I)}(k_n)~\tau_{\pi\Sigma-\pi\Sigma}^{(I)}(W-E_\alpha(\bm{p}_\alpha)-E_{N}(q_{n}))
 ~(u^{2}_{\nu,2I}(q,W-E_{\alpha})-u^{3}_{\nu,3I}(q_{n},W-E_{\alpha})) \nonumber\\
 &+\langle \pi\Sigma n|[[\pi\otimes N]_{I}\otimes \Sigma]_\Gamma\rangle~
 g_{\pi{N}}^{(I)}(k_\Sigma)~\tau_{\pi{N}-\pi{N}}^{(I)}(W-E_{\alpha}(p_{\alpha})-E_\Sigma(\bm{q}_\Sigma))
 ~(u^{2}_{\nu,3I}(q,W-E_{\alpha})-u^{3}_{\nu,2I}(q_{\Sigma},W-E_{\alpha})) \nonumber\\
 & +\langle \pi\Sigma n|[[\Sigma\otimes N]_{I}\otimes \pi]_\Gamma\rangle~
 g_{\Sigma{N}}^{(I)}(k_\pi)~\tau_{\Sigma{N}-\Sigma{N}}^{(I)}(W-E_{\alpha}(p_{\alpha})-E_\pi(\bm{q}_\pi))
 (u^{2}_{\nu,1I}(q,W-E_{\alpha})-(-1)^{I+\frac{1}{2}}u^{3}_{\nu,1I}(q_{\pi},W-E_{\alpha}))
 \Big].\nonumber\\
\label{eeq8}
\end{align}
\end{widetext}

Here, \(\mathcal{A}\) denotes the four-body transition amplitude from 
the \(K^- + {}^{6}\mathrm{Li}\) initial state to either the intermediate 
\(\alpha + (\bar{K}NN)_{s=1}\) configuration or directly to the 
\(\alpha + \pi\Sigma n\) final state. Since the explicit four-body dynamics 
is neglected in the present approach, the amplitude \(\mathcal{A}\) is set 
to unity. 
The quantity \(p_{\alpha}\) denotes the momentum of the spectator \(\alpha\) 
particle, while \(q_{i}\) represents the momentum of the spectator particle 
in the \(K^-np-\pi\Sigma{n}\) three-body system, when the neutron acts as the 
spectator particle. In Eq.~\ref{eeq8}, the functions \(u\) and \(\theta\) are 
obtained by solving the one-channel Faddeev--AGS equations.
\section{Two-body interactions}
\label{put}
A realistic description of the underlying two-body interactions is 
essential for studying the antikaon interaction with \(^{6}\mathrm{Li}\). 
In the present calculation, all two-body potentials are formulated in a 
separable form, which is particularly suitable for few-body scattering 
problems and allows for an efficient treatment of the three-body $\bar{K}NN$ 
subsystem. The antikaon--nucleon ($\bar{K}N$) interaction is primarily 
described by the SIDD1 and SIDD2 potentials, which are coupled-channel, 
energy-independent separable potentials constrained by low-energy $\bar{K}N$ 
scattering data and kaonic hydrogen measurements~\cite{p7}. The use of 
both potentials allows us to assess the sensitivity of the calculated 
observables to different realizations of the subthreshold $\bar{K}N$--$\pi\Sigma$ 
dynamics.

In addition to the phenomenological SIDD potentials, we employ an 
energy-dependent chiral $\bar{K}N$ potential derived from SU(3) chiral 
effective field theory~\cite{n99}. This model incorporates coupled-channel 
dynamics self-consistently. Due to its energy dependence, it generally 
yields a somewhat weaker attraction in the $\bar{K}N$ channel below 
threshold compared to energy-independent phenomenological potentials, 
which affects the binding mechanism and structure of the $\bar{K}NN$ 
subsystem. For SIDD1, the form factor is of Yamaguchi type,
\begin{equation}
g_{i}(p) = \frac{1}{p^2 + \beta_i^2},
\qquad i,j \in \{\bar{K}N, \pi\Sigma\},
\label{eeq10}
\end{equation} 
while for SIDD2 it takes the form
\begin{equation}
g_{i}(p) = \frac{1}{p^2 + \beta_i^2}
+ \frac{s \beta_i^2}{(p^2 + \beta_i^2)^2},
\qquad i,j \in \{\bar{K}N, \pi\Sigma\},
\label{eeq11}
\end{equation}
with the $s$-parameter in the $\bar{K}N$ channel set to zero. The main 
difference between SIDD1 and SIDD2 lies in the assumed pole structure of 
the $\Lambda(1405)$ resonance, corresponding to single-pole and two-pole 
scenarios, respectively. In this work, the $\pi N$ channel is neglected 
due to its minor impact on the low-energy dynamics.

The nucleon--nucleon (\(NN\)) interaction is modeled by the separable PEST 
potential, which reproduces low-energy \(NN\) scattering phase shifts and 
deuteron properties \cite{nn9}. This choice ensures a consistent treatment 
of the \(NN\) subsystem within the three-body \(\bar{K}NN\) framework, while 
the \(\alpha\) particle is treated as a spectator in the 
\(K^- + {}^{6}\mathrm{Li}\) reaction.

The hyperon--nucleon interaction in the \(\Sigma N\) channel is also included 
to account for possible final-state effects in the \(\pi\Sigma n\) system. 
In the isospin \(I = 1/2\) channel, the \(\Sigma N\) interaction is coupled 
with the \(\Lambda N\) channel and implemented in a rank-one separable form,
\begin{equation}
V_{ij}(p,p') = \lambda_{ij}\, g_i(p)\, g_j(p'),
\qquad i,j \in \{\Sigma N, \Lambda N\}.
\label{eeq12}
\end{equation}
where the parameters of the \(\Sigma N\) potential are given in Ref.~\cite{t1,t2}. 
All two-body interactions are restricted to the \(s\)-wave channel, consistent 
with the treatment of the \(\bar{K}N\) and \(NN\) subsystems. In the present 
calculations, the (\(\pi N\)) interaction is neglected, as its contribution to 
the low-energy dynamics of the system is expected to be minor.

For separable potentials, the corresponding two-body \(t\)-matrices are obtained 
analytically from the Lippmann--Schwinger equation,
\begin{equation}
t_{ij}(E) = V_{ij} + V_{ij} G_0(E) t_{ij}(E),
\label{eeq13}
\end{equation}
where \(G_0(E)\) is the free two-body Green's function. Owing to the separable 
structure of the potential, the solution can be expressed as
\begin{equation}
t_{ij}(p,p';E) = g_i(p)\, \tau_{ij}(E)\, g_j(p'),
\end{equation}
with the reduced propagator
\begin{equation}
\mathbf{\tau}(E) = \left[ \mathbf{1-\Lambda}\mathbf{G}(E) \right]^{-1} \mathbf{\Lambda},
\label{eeq14}
\end{equation}
where \(\mathbf{\Lambda}\) is the matrix of interaction strengths and 
\(\mathbf{G}(E)\) is the matrix of two-body Green’s functions. These 
two-body \(t\)-matrices serve as input for the three-body dynamics of the 
\(\bar{K}NN\) subsystem.
\section{Results and Discussions}
\label{result}
Before presenting the numerical results, we comment on the treatment of 
singularities in the integral equations and transition amplitudes. In this 
study, the \emph{Point Method}~\cite{q1,q2} is employed to handle the moving 
singularities arising from intermediate \(\bar{K}N\) and \(\pi\Sigma\) 
propagators. This approach discretizes the integral equations over carefully 
chosen momentum-space points, allowing for a stable and well-defined numerical 
evaluation of the amplitudes across the entire kinematical region. The Point 
Method ensures a consistent treatment of threshold effects and coupled-channel 
dynamics, yielding invariant-mass spectra free from spurious numerical artifacts.
\subsection{Stopped \(K^-\) absorption on \(^{6}\mathrm{Li}\)}

When an antikaon enters a material, it loses energy predominantly 
through electromagnetic interactions and may reach a low-energy regime 
in which its interaction with nuclei is governed by the strong force. 
In this domain, antikaon--nucleus dynamics provides direct access to 
subthreshold antikaon--nucleon interactions. 

In this section, we focus on antikaons in the near-threshold regime 
interacting with $^{6}\mathrm{Li}$ nuclei, which effectively corresponds 
to stopped or quasi-stopped $K^-$ absorption in the nuclear medium. The 
nuclear structure of \(^{6}\mathrm{Li}\) is described within a 
\(d\)--\(\alpha\) cluster approximation, treating it as a bound system 
of a deuteron and an \(\alpha\) particle. Absorption of a low-energy 
antikaon by the nucleus can populate several \(\pi\Sigma\) final states. 
The most relevant reaction channels, together with their estimated 
phase-space fractions and \(Q\)-values, are
\begin{widetext}
\begin{equation}
\begin{aligned}
& (1): K^{-}+ \,^{6}\mathrm{Li} \rightarrow (\pi+\Sigma)^{0}+n+\alpha,
\quad \sim 52\%, \quad Q = 94.3~\mathrm{MeV}, \\
& (2): K^{-}+ \,^{6}\mathrm{Li} \rightarrow (\pi+\Sigma)^{0}+d+t,
\quad \hspace{0.15cm} \sim 46\%, \quad Q = 76.7~\mathrm{MeV}.
\end{aligned}
\label{eeq15}
\end{equation}
\end{widetext}

\begin{figure}[h]
\centering
\vspace{0cm}
\hspace{-0cm}
\includegraphics[width=12cm]{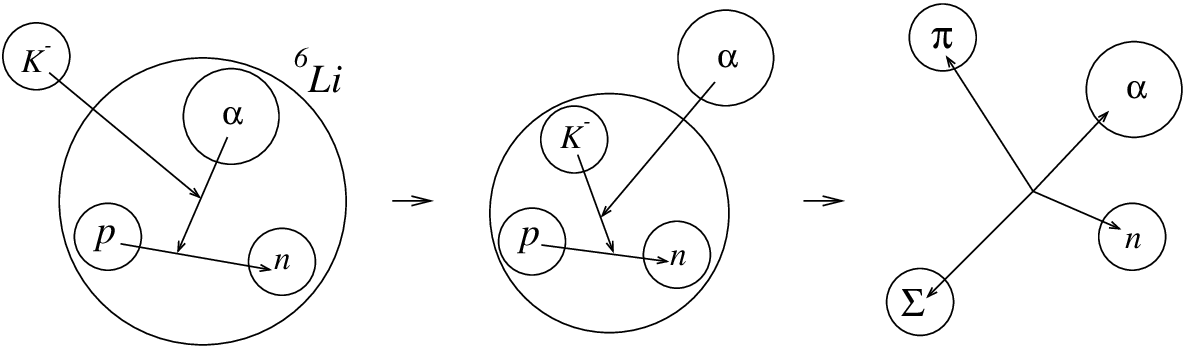} 
\vspace{-0cm}
\caption{(Color online) Schematic representation of the transition from 
\( K^{-} + \mathrm{^{6}Li} \) to \( \alpha + (\pi\Sigma n) \) via the intermediate 
\( \alpha + (\bar{K}NN)_{s=1} \) state.}
\label{fig1}
\end{figure}

The available phase space for each reaction scales as \(Q^{(3\mathcal{N}-5)/2}\), 
where \(\mathcal{N}\) denotes the number of particles in the final state~\cite{n4}. 
Reaction channels with a larger number of final particles are therefore suppressed, 
and the first two channels dominate, as shown in Fig.~\ref{fig1}. In the leading 
channel, the antikaon interacts primarily with the deuteron cluster, producing a 
\(\pi + \Sigma + n + \alpha\) final state.

Since this channel constitutes one of the dominant absorption mechanisms in 
low-energy \(K^-\) interactions with \(^{6}\mathrm{Li}\), the present work 
is devoted to its detailed investigation. In particular, we focus on the 
reconstruction of the \(\alpha\)-particle missing-mass spectrum. For the 
reaction \(K^- + {}^{6}\mathrm{Li} \rightarrow \pi + \Sigma + n + \alpha\), 
energy--momentum conservation implies that the missing mass associated with 
the detected \(\alpha\) particle is equivalent to the invariant mass of the 
\(\pi\Sigma n\) system. Consequently, the \(\alpha\)-particle spectrum provides 
direct access to the dynamics of the \(\pi\Sigma n\) final state and serves 
as a sensitive probe of the underlying antikaon--nucleon interaction, including 
possible resonance structures in the \(\pi\Sigma\) channel.

\begin{figure}[h]
\centering
\vspace{1cm}
\hspace{-0.cm}
\includegraphics[width=12.cm]{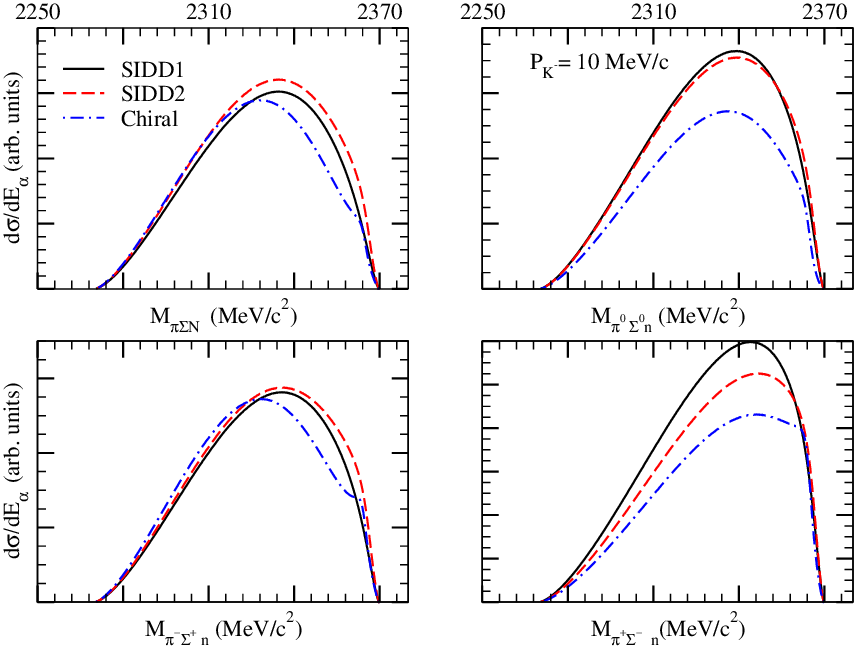} 
\vspace{-0.cm}
\caption{(Color online) 
The \((\pi\Sigma)^{0}n\) invariant mass spectrum for stopped 
kaon absorption on \(K^- + {}^{6}\mathrm{Li} \). 
}
\label{fig2}
\end{figure}

The results for the low-energy $K^-$–$^{6}\mathrm{Li}$ interaction at an 
incident kaon momentum of $10~\mathrm{MeV}/c$ are shown in Fig.~\ref{fig2}. 
Fig.~\ref{fig2} displays the $\pi\Sigma n$ invariant-mass spectrum and the 
corresponding $\alpha$-particle missing-mass distribution, highlighting the 
dominant absorption mechanism and the manifestation of the $\Lambda(1405)$ 
resonance. The chosen kaon momentum corresponds to the low-momentum antikaons 
produced via $\phi$-meson decay at the DA$\Phi$NE electron--positron collider, 
making it experimentally accessible. After being produced, the $K^-$ mesons 
lose their kinetic energy through electromagnetic interactions in the target 
material and are subsequently absorbed by the nucleus in the near-threshold or 
quasi-stopped regime. As a result, the reaction dynamics is dominated by 
$s$-wave $\bar{K}N$ interactions and is particularly sensitive to subthreshold 
effects in the $\bar{K}N$ system. Moreover, the effective low-energy nature of 
the absorption process reduces kinematical broadening and enhances the visibility 
of resonance structures in the $\pi\Sigma n$ invariant-mass distribution, thereby 
improving the resolution of the $\alpha$-particle missing-mass spectrum and 
facilitating a direct comparison with experimental observables.
\subsection{In-flight kaon--lithium reaction}
\begin{figure*}[htb]
\centering
\vspace{1cm}
\hspace{-0.cm}
\includegraphics[width=8.3cm]{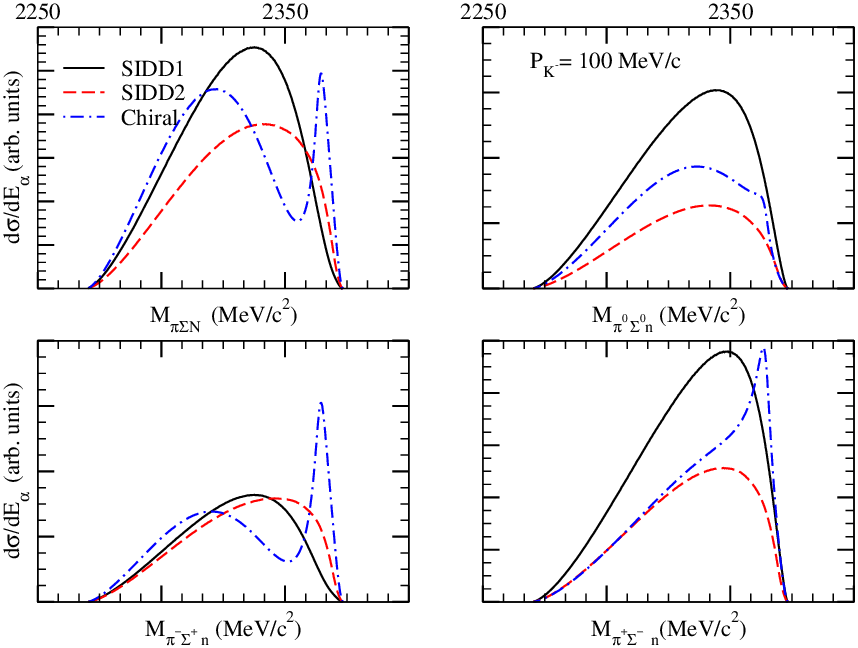} 
\hspace{0.7cm}
\includegraphics[width=8.3cm]{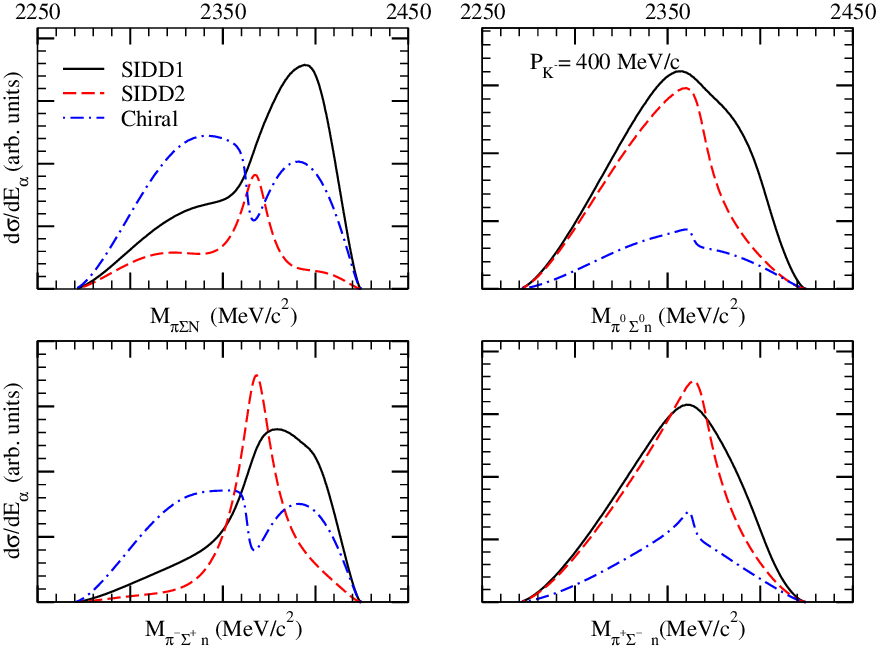} 
\vspace{-0.cm}
\caption{(Color online) 
The invariant mass spectrum of the \((\pi\Sigma)^{0}n\) system, equivalently 
expressed as the \(\alpha\)-particle missing-mass spectrum, for the transition
\(K^- + {}^{6}\mathrm{Li} \to \alpha + (\pi\Sigma)^{0} n\). The kaon momentum 
in the center-of-mass frame is \(100\) and \(400\,\mathrm{MeV}/c\).
}
\label{fig3}
\end{figure*}

\begin{figure*}[htb]
\centering
\vspace{1.2cm}
\hspace{-0cm}
\includegraphics[width=8.3cm]{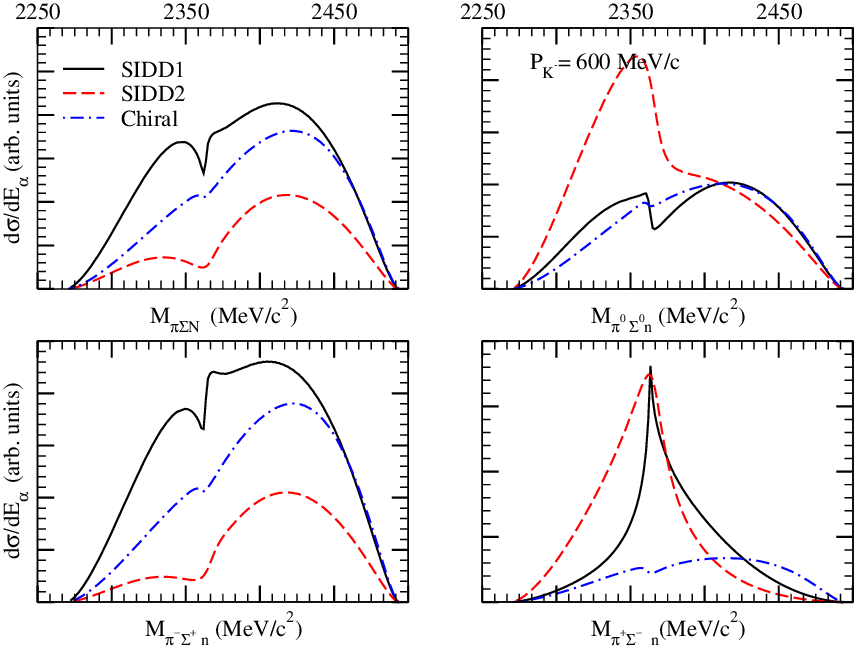} 
\hspace{0.7cm}
\includegraphics[width=8.3cm]{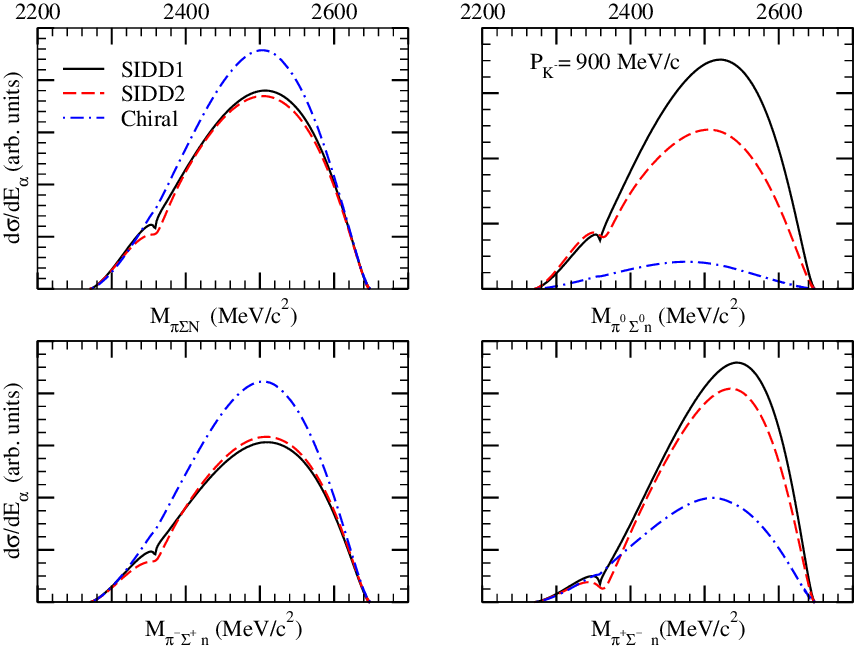} 
\vspace{-0.cm}
\caption{(Color online) The descriptions follow those given in Fig.~\ref{fig2}; 
however, in the present case the kaon momentum in the center-of-mass system is 
fixed at \(600\) and \(900\,\mathrm{MeV}/c\).
}
\label{fig4}
\end{figure*}

In this section, we investigate the invariant-mass spectrum of the \(\pi\Sigma n\) 
system for in-flight kaon--lithium reactions, employing different models of the 
antikaon--nucleon interaction and several values of the incident kaon momentum. 
The calculations are performed for kaon momenta of \(100\), \(400\), \(600\), and 
\(900\,\mathrm{MeV}/c\), spanning a wide kinematical range from near-threshold 
energies up to the region well above the \(\bar{K}N\) threshold. This momentum 
range allows for a systematic study of the evolution of the reaction dynamics and 
of the stability of the extracted spectral features against changes in the incident 
kaon energy. The calculated spectra are shown in Figs.~\ref{fig3} and \ref{fig4}.

For all considered kaon momenta and interaction models, a pronounced enhancement 
associated with the \(\Lambda(1405)\) resonance is observed in the \(\pi\Sigma n\) 
invariant-mass spectra. The persistence of this structure over a broad momentum 
range indicates that the formation mechanism of the \(\Lambda(1405)\) in the present 
reaction remains effective even for in-flight kaons and is not strongly suppressed 
at higher incident momenta. At the same time, the detailed shape and peak position 
of the spectral strength exhibit a noticeable dependence on the underlying \(\bar{K}N\) 
interaction model, reflecting the sensitivity of the \(\pi\Sigma n\) system to the 
subthreshold behavior of the antikaon--nucleon interaction.

To enable a more realistic description and a direct comparison with experimental 
data, the invariant-mass spectra are evaluated in the physical particle basis 
rather than in the isospin basis. Explicit calculations are carried out for the 
\(\pi^-\Sigma^+ n\), \(\pi^+\Sigma^- n\), and \(\pi^0\Sigma^0 n\) channels. This 
treatment properly incorporates isospin-breaking effects arising from physical 
mass differences and channel-dependent kinematics, allowing for a meaningful 
comparison of individual channel contributions and line shapes with experimentally 
measured spectra.

The combined analysis of different incident kaon momenta, interaction models, 
and particle channels provides a comprehensive picture of the \(\pi\Sigma n\) 
invariant-mass distribution in in-flight kaon-induced reactions. In particular, 
the robust appearance of the \(\Lambda(1405)\) signal across all considered 
scenarios underscores the suitability of this reaction as a sensitive probe of 
\(\bar{K}N\) interaction and of the resonance dynamics in the \(\pi\Sigma\) 
channel.

At energies above the $\bar{K}N$ threshold, the $\pi\Sigma$ invariant-mass 
spectrum is no longer dominated solely by the subthreshold structure of the 
$\Lambda(1405)$, but reflects the coupled-channel $\bar{K}N$--$\pi\Sigma$ 
dynamics within the $s$-wave sector. In this energy region, the spectrum 
becomes sensitive to the opening of the $\bar{K}N$ channel, nonresonant 
$s$-wave contributions, and the energy dependence and phase motion of the 
scattering amplitude. Consequently, the behavior of the $\pi\Sigma$ spectrum 
above threshold provides important constraints on the strength and energy 
dependence of the $s$-wave $\bar{K}N \leftrightarrow \pi\Sigma$ coupling. 
These effects become particularly visible at higher incident kaon momenta, 
such as $600$ and $900\,\mathrm{MeV}/c$, as illustrated in Fig.~\ref{fig4}.

In the present calculation, all two-body interactions are therefore restricted 
to the $s$-wave channel. Accordingly, even in the energy region above the 
$\bar{K}N$ threshold, the $\pi\Sigma$ invariant-mass spectrum reflects 
exclusively the $s$-wave dynamics of the coupled $\bar{K}N$--$\pi\Sigma$ 
system. While contributions from higher partial waves may start to become 
relevant at sufficiently high energies, such effects are not included in 
the present approach. The inclusion of higher partial waves in the $\bar{K}N$ 
interaction, as discussed for example in Ref.~\cite{fe1}, is beyond the 
scope of the present work. This restriction is motivated by the fact that 
the $\Lambda(1405)$ resonance is known to be generated predominantly 
through coupled-channel $s$-wave dynamics. The resulting $\pi\Sigma$ 
spectrum thus provides direct information on the energy dependence and 
phase motion of the $s$-wave $\bar{K}N \leftrightarrow \pi\Sigma$ amplitude 
and on the underlying coupled-channel mechanism responsible for the formation 
of the $\Lambda(1405)$.

Recently, a fully three-coupled-channel calculation including the $\bar{K}N$, 
$\pi\Sigma$, and $\pi\Lambda$ channels has been reported by Shevchenko \cite{sh5}, 
showing improved agreement with the J-PARC E15 data and modified binding energy 
and width of the $K^- d$ quasi-bound state. In the present work, the $\pi\Lambda$ 
channel is effectively taken into account through the employed $\bar{K}N$ interaction 
models, while the $\bar{K}N$--$\pi\Sigma$ dynamics is treated explicitly within the 
AGS framework. Although an explicit inclusion of the $\pi\Lambda$ channel may 
quantitatively affect the extracted widths and fine spectral structures, the 
qualitative behavior of the observables shown in Figs.~\ref{fig2}--\ref{fig4} is 
expected to remain stable. A systematic extension of the present AGS calculations 
to a fully coupled-channel treatment constitutes an important subject for future 
investigations.

The results presented in Figs.~\ref{fig2}, \ref{fig3}, and \ref{fig4} allow a 
systematic investigation of the sensitivity of the $K^- + {}^{6}\mathrm{Li}$ reaction 
observables to the underlying pole structure of the $\Lambda(1405)$. The one-pole 
and two-pole $\bar{K}N$ interaction models employed in this work lead to characteristic 
differences in the shape and strength of the calculated spectra, reflecting the 
distinct analytic structures of the scattering amplitudes. While these differences 
do not necessarily correspond directly to experimentally observed peak positions, 
they provide valuable information on how the $\Lambda(1405)$ dynamics is embedded 
in a realistic four-body reaction process.

At present, the aim of this study is not a direct extraction of the $\Lambda(1405)$ 
pole structure from experimental data, but rather a theoretical assessment of model 
dependence within a unified AGS framework. A detailed comparison with experimental 
spectra, such as those from the J-PARC E15 experiment, requires additional elements, 
including a consistent treatment of background contributions and detector effects. 
Such an extension is beyond the scope of the present work and will be addressed in 
future studies.
\section{Conclusions}
\label{conc}
We have investigated the low- and intermediate-energy interaction 
of antikaons with the \(^{6}\mathrm{Li}\) nucleus through the 
\(\pi\Sigma n\) invariant-mass spectrum. The nucleus was described 
within a \(d\)--\(\alpha\) cluster approximation, reducing the 
original four-body problem to an effective three-body \(\bar{K}NN\) 
system while retaining the dominant absorption mechanism.

The antikaon--nucleon interaction was modeled using the SIDD1, 
SIDD2 and chiral potentials, combined with the PEST nucleon--nucleon 
interaction, with all two-body forces restricted to the \(s\)-wave 
channel. Calculations were performed for incident kaon momenta 
of \(100\), \(400\), \(600\), and \(900\,\mathrm{MeV}/c\), covering 
a wide kinematical range from near threshold to well above the 
\(\bar{K}N\) threshold.

A pronounced structure associated with the \(\Lambda(1405)\) resonance 
is predicted in the \(\pi\Sigma n\) invariant-mass spectra for all 
considered momenta. The persistence of this signal demonstrates that the 
reaction \(K^- + {}^{6}\mathrm{Li} \to \alpha + \pi + \Sigma + n\) can 
provide a robust probe of subthreshold \(\bar{K}N\) dynamics. At the same 
time, noticeable differences between the SIDD1, SIDD2 and chiral results 
indicate a clear sensitivity to the underlying \(\bar{K}N\) interaction 
and the pole structure of the \(\Lambda(1405)\).

The analysis was carried out explicitly in the physical particle channels 
\(\pi^-\Sigma^+ n\), \(\pi^+\Sigma^- n\), and \(\pi^0\Sigma^0 n\), allowing 
a prediction of the features that could be observed in future experiments. 
The corresponding \(\alpha\)-particle missing-mass spectra provide 
complementary information on the \(\pi\Sigma n\) invariant-mass distribution.

Within the present \(s\)-wave framework, our results indicate that 
kaon-induced reactions on \(^{6}\mathrm{Li}\) are well suited for 
studying the antikaon--nucleon interaction and the formation mechanism 
of the \(\Lambda(1405)\). Extensions including higher partial waves and 
more refined nuclear dynamics will be necessary to achieve a quantitative 
description at higher incident kaon momenta.

\end{document}